\documentclass{appolb}
\usepackage{epsfig}
\begin{document}
% \eqsec  % uncomment this line to get equations numbered by (sec.num)
\title{Lifetime of metastable states and suppression
of noise in Interdisciplinary Physical Models
\thanks{Presented at the $19^{th}$ Marian Smoluchowski Symposium on Statistical Physics,
Krak\'ow, Poland, May 14-17, 2006 }
% you can use '\\' to break lines
}
\author{B. Spagnolo$^a$\footnote{e-mail: spagnolo@unipa.it},
A. A. Dubkov$^b$, A. L. Pankratov$^c$, \\ E. V. Pankratova$^d$, A.
Fiasconaro$^{a,e,f}$, A. Ochab-Marcinek$^e$
\address{$^a$Dipartimento di Fisica e Tecnologie Relative, Universit\`a di
Palermo\\
CNISM - Unit\`a di Palermo, Group of Interdisciplinary
Physics\footnote{http://gip.dft.unipa.it} \\
Viale delle Scienze, I-90128 Palermo, Italy\\
$^b$Radiophysics Department, Nizhny Novgorod State University, 23
Gagarin ave., 603950 Nizhny Novgorod,
Russia\\
$^c$Institute for Physics of Microstructures of RAS, GSP-105, Nizhny
Novgorod, 603950, Russia\\
$^d$Mathematical Department, Volga State Academy, Nesterov street 5,
\\ Nizhny Novgorod, 603600, Russia\\
$^e$Marian Smoluchowski Institute of Physics, Jagellonian University\\
Reymonta 4, 30–059 Kraków, Poland\\
$^f$Mark Kac Center for Complex Systems Research, Jagellonian University\\
Reymonta 4, 30–059 Kraków, Poland}}

 \maketitle
\begin{abstract}
Transient properties of different physical systems with metastable
states perturbed by external white noise have been investigated. Two
noise-induced phenomena, namely the noise enhanced stability and the
resonant activation, are theoretically predicted in a piece-wise
linear fluctuating potential with a metastable state. The
enhancement of the lifetime of metastable states due to the noise,
and the suppression of noise through resonant activation phenomenon
will be reviewed in models of interdisciplinary physics: (i)
dynamics of an overdamped Josephson junction; (ii) transient regime
of the noisy FitzHugh-Nagumo model; (iii) population dynamics.

\end{abstract}

%Keywords: Statistical Mechanics, Population Dynamics,
%Noise-induced effects


% 05.10.-a  Computational methods in statistical physics and nonlinear
% dynamics
% 05.40.-a Fluctuation phenomena, random processes, noise, and Brownian motion
% 87.23.Cc Population dynamics and ecological pattern formation

\section{Introduction}

Metastability is a generic feature of many nonlinear systems, and
the problem of the lifetime of metastable states involves
fundamental aspects of nonequilibrium statistical mechanics.
Nonequilibrium systems are usually open systems which strongly
interact with environment through exchanging materials and energy,
which can be modelled as noise. The investigation of noise-induced
phenomena in far from equilibrium systems is one of the approaches
used to understand the behaviour of physical and biological complex
systems. Specifically the relaxation in many natural complex systems
proceeds through metastable states and this transient behaviour is
observed in condensed matter physics and in different other fields,
such as cosmology, chemical kinetics, biology and high energy
physics~\cite{gun83}-\cite{gle05}. In spite of such ubiquity, the
microscopic understanding of metastability still raises fundamental
questions, such as those related to the fluctuation-dissipation
theorem in transient dynamics~\cite{par05}.

Recently, the investigation of the thermal acivated escape in
systems with fluctuating metastable states has led to the discovery
of resonancelike phenomena, characterized by a nonmonotonic behavior
of the lifetime of the metastable state as a function of the noise
intensity or the driving frequency. Among these we recall two of
them, namely the resonant activation (RA)
phenomenon~\cite{doe92}-\cite{ale06}, whose signature is a minimum
of the lifetime of the metastable state as a function of a driving
frequency, and the noise enhanced stability (NES)~\cite{man96},
\cite{dub04}-\cite{spa04}. This resonancelike effect, which
contradicts the monotonic behavior predicted by the Kramers
formula~\cite{kra40,han90}, shows that the noise can modify the
stability of the system by enhancing the lifetime of the metastable
state with respect to the deterministic decay time. Specifically
when a Brownian particle is moving in a potential profile with a
fluctuating metastable state, the NES effect is always obtained,
regardless of the unstable initial position of the particle. Two
different dynamical regimes occur. These are characterized by: (i) a
monotonic behavior with a divergence of the lifetime of the
metastable state when the noise intensity tends to zero, for a given
range of unstable initial conditions (see for detail
Ref.~\cite{fia05}), which means that the Brownian particle will be
trapped into the metastable state in the limit of very small noise
intensities; (ii) a nonmonotonic behavior of the lifetime of the
metastable state as a function of noise intensity. The noise
enhanced stability effect implies that, under the action of additive
noise, a system remains in the metastable state for a longer time
than in the deterministic case, and the escape time has a maximum as
a function of noise intensity. We can lengthen or shorten the mean
lifetime of the metastable state of our physical system, by acting
on the white noise intensity. The noise-induced stabilization, the
noise induced slowing down in a periodical potential, the noise
induced order in one-dimensional map of the Belousov-Zhabotinsky
reaction, and the transient properties of a bistable kinetic system
driven by two correlated noises, are akin to the NES phenomenon
\cite{hir82}.

In this paper we will review these two noise-induced effects in
models of interdisciplinary physics, ranging from condensed matter
physics to biophysics.  Specifically in the first section, after
shortly reviewing the theoretical results obtained with a model
based on a piece-wise linear fluctuating potential with a metastable
state, we focus on the noise-induced effects RA and NES. In the next
sections, we show how the enhancement of the lifetime of metastable
states due to the noise and the suppression of noise, through
resonant activation phenomenon, occur in the following
interdisciplinary physics models: (i) dynamics of an overdamped
Josephson junction; (ii) transient regime of the noisy
FitzHugh-Nagumo model; (iii) population dynamics.

\section{The model}

\begin{figure}[ptb]
\centerline{\epsfxsize=2.5in\epsfbox{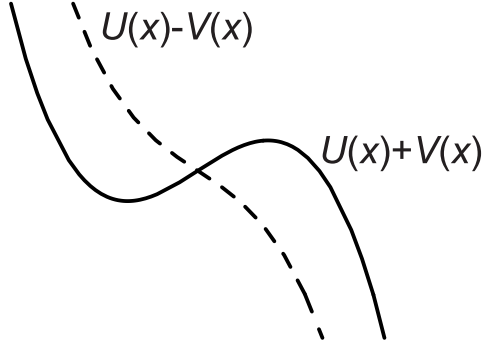}} \caption{Switching
potential with metastable state.} \label{fig1}
\end{figure}

As an archetypal model for systems with a metastable state and
strongly coupled with the noisy environment, we consider the
one-dimensional overdamped Brownian motion in a fluctuating
potential profile

\begin{equation}
\frac{dx}{dt}  = -\frac{\partial \left[U\left(x\right)  +
V\left(x\right) \eta
 \left(  t\right) \right]}{\partial x}
+ \xi(t) ,
 \label{Lang}%
\end{equation}
where $x(t)$ is the displacement of the Brownian particle and
$\xi(t)$ is the white Gaussian noise with the usual statistical
properties: $<\xi(t)>\, = 0$,  $<\xi(t) \xi(t')> =\, 2 D \;
\delta(t-t')$. The variable $\eta\left( t\right)  $ is the Markovian
dichotomous noise, which takes the values $\pm1$ with the mean
flipping rate $\nu$. The potential profile $U\left( x\right)
+V\left(x\right)  $ corresponds to a metastable state, and $U\left(
x\right) - V\left( x\right)$ corresponds to an unstable one, with a
reflecting boundary at $x\rightarrow-\infty$ and an absorbing
boundary at $x\rightarrow +\infty$ (see Fig.~\ref{fig1}). Starting
from the well-known expression for the probability density of the
process $x(t)$

\begin{equation}
P\left( x,t\right) = \langle\delta\left(x-x(t)\right)\rangle
\label{known}
\end{equation}
and using the auxiliary function $Q\left(x,t\right)$
\begin{equation}
Q\left(  x,t\right)  =\langle\eta\left(  t\right)  \delta\left(
x-x(t)\right)  \rangle,
\label{qu}
\end{equation}
and the Eq.~(\ref{Lang}), we obtain the set of differential
equations

\begin{eqnarray}
\frac{\partial P(x,t)}{\partial t} &=& \frac{\partial}{\partial
x}\left[ U^{\prime}\left(x\right)  P +V^{\prime}\left( x\right)
Q\right] + D\frac{\partial^{2}P}{\partial x^{2}},\nonumber\\
\frac{\partial Q(x,t)}{\partial t} &=& -2\nu Q +
\frac{\partial}{\partial x}\left[ U^{\prime}\left(x\right) Q +
V^{\prime}\left(x\right) P\right] + D\frac{\partial^{2}Q}{\partial
x^{2}}. \label{two}
\end{eqnarray}
The average lifetime of Brownian particles in the interval $\left(
L_{1},L_{2}\right)$, with the initial conditions $P\left( x,0\right)
= \delta\left(x-x_{0}\right)$ and $Q\left(x,0\right) =\pm
\delta\left(x-x_{0}\right)$, is
\begin{equation}
\tau\left(  x_{0}\right) =\int_{0}^{\infty}dt\int_{L_{1}}^{L_{2}} P
\left(x,t\left\vert x_{0},0\right.  \right)  dx =
\int_{L_{1}}^{L_{2}}Y\left( x,x_{0},0\right)  dx \;,
\label{taudef}
\end{equation}
where $Y\left( x,x_{0},s\right)$ is the Laplace transform of the
conditional probability density $P \left(x,t\left\vert
x_{0},0\right. \right)$.
After Laplace transforming Eqs.~(\ref{two}), with above-mentioned
initial conditions and using the method proposed in
Refs.~\cite{mal96,agu99,mal97}, we can express the lifetime
$\tau(x_{0})$ as

\begin{equation}
\tau\left(  x_{0}\right)  =\int_{L_{1}}^{L_{2}}Z_{1}\left(
x,x_{0}\right) dx,
\label{tau}
\end{equation}
where $Z_1(x,x_0)$ is the linear coefficient of the expansion of the
function $s Y(x,x_0,s)$ in a power series in $s$. By Laplace
transforming the auxiliary function $Q(x,t)$ in $R(x,x_{0},s)$, and
expanding the function $s R(x,x_{0},s)$ in similar power series, we
obtain the following closed set of integro-differential equations
for the functions $Z_{1}\left(x,x_{0}\right)$ and
$R_{1}\left(x,x_{0}\right)$~\cite{dub04}

\begin{eqnarray}
DZ_{1}^{\prime}+U^{\prime}\left(  x\right)  Z_{1}+V^{\prime}\left(
x\right)
R_{1} =-\theta\left(  x-x_{0}\right) ,\nonumber\\
DR_{1}^{\prime}+U^{\prime}R_{1}+V^{\prime}Z_{1} =2\nu\int_{-\infty}%
^{x}R_{1}dy\mp\theta\left(  x-x_{0}\right)  ,
\label{finally}%
\end{eqnarray}
where $\theta(x-x_{0})$ is the Heaviside step function, and
$R_{1}\left(x,x_{0}\right)$ is the linear coefficient of the
expansion of the function $s R(x,x_0,s)$. We put equal to zero the
probability flow at the reflecting boundary $x = - \infty$. These
general equations (\ref{tau}) and (\ref{finally}) allow to calculate
the average lifetime for potential profiles with metastable states.
We may consider two mean lifetimes $\tau_{+}\left( x_{0}\right)  $
and $\tau _{-}\left( x_{0}\right) $, depending on the initial
configuration of the randomly switching potential profile: $U\left(
x\right) + V\left(x\right)$ or $U\left( x\right) -V\left( x\right)$.
The average lifetime (\ref{tau}) is equal to $\tau_{+}\left(
x_{0}\right)$, when we take the sign "$-$" in the second equation of
system (\ref{finally}), and vice versa for
$\tau_{-}\left(x_{0}\right)$.

We consider now the following piece-wise linear potential profile

\begin{eqnarray}
U(x)=\left\{
\begin{array}
[c]{ll} + \infty, & x<0\\
0, & 0\leq x\leq L\\
k\left(  L-x\right)  , & x>L
\end{array}
\right.
\label{piecewise}
\end{eqnarray}
and $V\left(x\right) = ax$ ($x>0$, $0<a<k$). Here we consider the
interval $L_1 = 0, L_2 = b$, with $b > L$. After solving the
differential equations (\ref{finally}) with the potential profile
(\ref{piecewise}), and by choosing the initial position of Brownian
particles at $x = 0$, we get the exact mean lifetime

\begin{equation}
\tau_{-}\left(x_{in} = 0\right)  =\frac{b}{k}+\frac{\nu
L^{2}}{\Gamma^{2}} + \frac{a}{2\nu\Gamma^{4}}\;f(D,\Gamma,\nu),
\label{main}
\end{equation}
where $\Gamma = \sqrt{a^2 + 2\nu D}$. Here $f(D,\Gamma,\nu)$, which
is also a function of the potential parameters $a,b,L,k$, has a
complicated expression in terms of parameters $D, \Gamma$ and
$\nu$~\cite{dub04}. It is worthwhile to note that Eq.~(\ref{main})
was derived without any assumptions on the white noise intensity $D$
and on the mean rate of flippings $\nu$ of the potential.

To look for the NES effect, which is observable at very small noise
intensity~\cite{man96,agu01}, we derive the average life time in the
limit $D \rightarrow 0$

\begin{equation}
\tau_{-}\left(x_{in} = 0\right) = \tau_{0}+\frac{D}{a^{2}} \;
g(q,\omega,s)+o \left(D\right) \ ,
\label{simp}
\end{equation}
where $\omega=\nu L/k$, $q=a/k$ and $s=2\omega\left( b/L-1\right)
/\left( 1-q^{2}\right)$ are dimensionless parameters. Here

\begin{eqnarray}
g(q,\omega,s) &=&  \frac{3q^2+4q-5}{2(1-q^2)} + 2\omega
\frac{3q^2+q-3}{q(1-q^2)} - \frac {2\omega^2}{q^2} \nonumber \\
& & + se^{-s}\frac {q^3(1+q^2)}{(1+q)(1-q^2)}+ \left(
1-e^{-s}\right) \frac {q(1-q^2-2q^3)}{2(1-q^2)}
\label{grad}
\end{eqnarray}
and
\begin{equation}
\tau_{0}=\frac{2L}{a}+\frac{\nu L^{2}}{a^{2}}+\frac{b-L}{k}-\frac{q(1-q)}%
{2\nu}\left(  1-e^{-s}\right)
\label{tau0}
\end{equation}
is the mean lifetime in the absence of white Gaussian noise $(D=0)$.
The condition to observe the NES effect can be expressed by the
inequality

\begin{equation}
g\left(q,\omega,s\right)>0.
\label{NES}
\end{equation}
The main conclusions from the analysis of the inequality~(\ref{NES})
are: (i) the NES effect occurs at $q\simeq1$, i.e. at very small
steepness $k-a=k\left( 1-q\right) $ of the reverse potential barrier
for the metastable state: for this potential profile, a small noise
intensity can return particles into potential well, after they
crossed the point $L$; (ii) for a fixed mean flipping rate, the NES
effect increases when $q \rightarrow 1$, and (iii) for fixed
parameter $q$ the effect increases when $\omega\rightarrow0$,
because Brownian particles have enough time to move back into
potential well.

\begin{figure}[ptb]
\centerline{\epsfxsize=2.5in\epsfbox{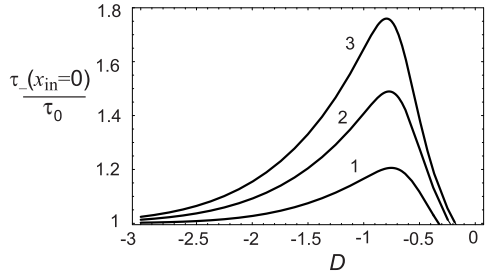}}
\caption{Semilogarithmic plot of the normalized mean lifetime
$\tau_{-}\left(x_{in} = 0\right) /\tau_{0}$ vs the white noise
intensity $D$ for three values of the dimensionless mean flipping
rate $\omega=\nu L/k$: $0.03$ (curve $1$), $0.01$ (curve $2$),
$0.005$ (curve $3$). Parameters are $L=1$, $k=1$, $b=2$, and
$a=0.995$.} \label{fig2}
\end{figure}

\begin{figure}[ptb]
\centerline{\epsfxsize=2.5in\epsfbox{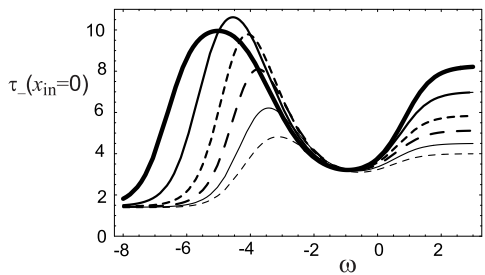}}
\caption{Semilogarithmic plot of the mean lifetime $\tau_{-}\left(
0\right)  $ vs the dimensionless mean flipping rate $\omega=\nu L/k$
for seven noise intensity values. Specifically from top to bottom on
the right side of the figure: $D = 0.08,0.1,0.13,0.16,0.2,0.25$. The
other parameters are the same as in Fig.~\ref{fig2}.} \label{fig3}
\end{figure}

Under very large noise intensity $D$, the Brownian particles
''\emph{do not see}'' the fine structure of potential profile and
move as in the fixed potential $U(x) = -kx$. Therefore the average
life time decreases with noise intensity, tending to the value $b/k$
as follows from Eq.~(\ref{main}) in the limit $D\rightarrow\infty$.
In Fig.~\ref{fig2} we show the plots of the normalized mean lifetime
$\tau _{-}\left(x_{in} = 0\right) /\tau_{0}$, Eq. (\ref{main}), as a
function of the noise intensity $D$ for three values of the
dimensionless mean flipping rate $\omega=\nu L/k$: $0.03$, $0.01$,
$0.005$. The values of the parameters of the potential profile are:
$L = k = 1, a = 0.995, b =2$.
The maximum value of the average lifetime and the range of noise
intensity values, where NES effect occurs, increases when $\omega$
decreases.
By using exact Eq.~(\ref{main}) we have also investigated the
behaviour of the mean lifetime $\tau_{-}(x_{in} = 0)$ as a function
of switchings mean rate $\nu$ of the potential profile. In
Fig.~\ref{fig3} we plot this behaviour for six values of the noise
intensity, namely: $D = 0.08,0.1,0.13,0.16,0.2,0.25$.
At very slow flippings $(\nu\rightarrow0)$ we obtain

\begin{equation}
\tau_{- (\nu\rightarrow0)}(x_{in} = 0)  \simeq\tau_{d} -
\frac{D\left( 1-e^{-aL/D} \right) }{a^{2}\left(  1+q\right)  },
\label{slow}%
\end{equation}
i.e. the average lifetime of the fixed unstable potential $U(x)
-ax$. Here $\tau_d = L/a + (b - L)/(k+a)$ is the deterministic time
at zero frequency ($\nu = 0$). While for very fast switchings
$(\nu\rightarrow\infty)$ we obtain
\begin{equation}
\tau_{- (\nu\rightarrow\infty)}(x_{in} = 0)  \simeq \frac{b}{k} +
\frac{L^{2}}{2D} , \label{fast}%
\end{equation}
i.e. the mean lifetime for average potential $U\left(  x\right) $.
All limiting values of $\tau_{-}(x_{in} = 0)$, expressed by
Eqs.~(\ref{slow}) and (\ref{fast}), are shown in Fig.~\ref{fig3}. At
intermediate rates the average escape time from the metastable state
exhibits a minimum at $\omega=0.1$, which is the signature of the
resonant activation (RA) phenomenon~\cite{doe92}-\cite{dyb02}. If
the potential fluctuations are very slow, the average escape time is
equal to the average of the crossing times over upper and lower
configurations of the barrier, and the slowest process determines
the value of the average escape time~\cite{doe92}. In the limit of
very fast fluctuations, the Brownian particle ''\emph{sees}'' the
average barrier and the average escape time is equal to the crossing
time over the average barrier. In the intermediate regime, the
crossing is strongly correlated with the potential fluctuations and
the average escape time exhibits a minimum at a resonant fluctuation
rate. Specifically, for $D\ll1$ and the parameter values of the
potential ($a = 0.995, L = k = 1, b = 2$), we obtain from
Eqs.~(\ref{slow}) and (\ref{fast}): $\tau_{-
(\nu\rightarrow0)}(x_{in} = 0) \simeq 1.5 - D/2$, and $\tau_{-
(\nu\rightarrow\infty)}(x_{in} = 0) \simeq 2 + 1/(2D)$, that is

\begin{equation}
\tau_{- (\nu\rightarrow0)}(x_{in} = 0)\ll \tau_{-
(\nu\rightarrow\infty)}(x_{in} = 0)
\end{equation}
which is consistent with the physical picture for which we have at
zero frequency of switchings the unstable initial configuration of
the potential (see Fig.~\ref{fig1}), and at very fast switchings
($\nu\rightarrow\infty$) the average configuration of the potential,
which in our case has not barrier. For $D\gg1$, because

\begin{equation}
\lim_{D\rightarrow\infty} D\left( 1-e^{-aL/D} \right) = aL \, ,
\end{equation}
we have $\tau_{- (\nu\rightarrow0)}(x_{in} = 0) = b/(k+a) \simeq 1$,
and $\tau_{- (\nu\rightarrow\infty)}(x_{in} = 0) = 2$. So, for the
noise intensity values used in our calculations shown in
Fig.~\ref{fig3}, ranging from $D = 0.08$ to $D = 0.25$, the limiting
values for the average lifetime are: $\tau_{-
(\nu\rightarrow0)}(x_{in} = 0) \simeq (1.38 \div 1.46)$, and
$\tau_{- (\nu\rightarrow\infty)}(x_{in} = 0) \simeq (4 \div 8)$,
which are consistent with the limiting values shown in
Fig.~\ref{fig3}, and evaluated directly from exact
expression~(\ref{main}).

Moreover, in Fig.~\ref{fig3} a \emph{new resonance-like behaviour},
is observed. The mean lifetime of the metastable state
$\tau_{-}(x_{in} = 0)$ exhibits a \emph{maximum}, between the slow
limit of potential fluctuations (static limit) and the RA minimum,
as a function of the mean fluctuation rate of the potential, . This
maximum occurs for a value of the barrier fluctuation rate on the
order of the inverse of the time $\tau_{up}(D)$ required to escape
from the metastable fixed configuration
\begin{equation}
\tau_{up}\left(  D\right)  = \frac{b-L}{k-a} - \frac{L}{a} +
\frac{D\left(
e^{aL/D}-1\right)  }{a^{2}\left(  1-q\right)  }.
\label{upper-tau}%
\end{equation}
Specifically we observe that this maximum increases with decreasing
noise intensity $D$ and at the same time the position of the maximum
is shifted towards lower values of the dimensionless mean flipping
rate $\omega$. In fact from Eq.~(\ref{upper-tau}) we have that the
average time required to escape from the metastable fixed
configuration $\tau_{up}$ increases, consequently the corresponding
rate of the barrier fluctuations $\omega_{max}\simeq 1/\tau_{up}(D)$
decreases, as shown in Fig.~\ref{fig3}. We can also estimate the
value of the maximum ($\tau_{-_{max}}(x_{in} = 0)$) and its position
($\omega_{max}$), by expanding Eqs.~(\ref{simp})-(\ref{tau0}) in a
power series up to the second order in $\omega$. Using the same
parameter values of the potential we have: $\omega = \nu$, $s =
2\omega /(1-q^2)$, $s e^{-s} \approx s - s^2$, and $1 -
e^{-s}\approx s - \frac{s^2}{2}$. We obtain finally:

\begin{equation}
\tau_{-}\left(x_{in} = 0\right) \approx 2.5 + (98.7)D + \omega[51 +
(347.4)D] - (2\times 10^6)\omega^2 D\;.
\label{tau_approx}
\end{equation}
For $D = 0.1$ we have: $\tau_{-_{max}}\approx 12.3$ and
$\omega_{max} \approx 2\times10^{-4}$, which are an estimate of the
coordinates of the maximum of the corresponding curve in
Fig.~\ref{fig3}. From small noise intensity $D\rightarrow0$, from
Eq.~(\ref{tau_approx}), we obtain: $\tau_{-_{max}}\approx \frac{0.6
\times 10^{-3}}{D} + O(D)$. The maximum of the average lifetime
$\tau_{-_{max}}$, therefore, increases when the noise intensity
decreases as shown in Fig.~\ref{fig3}.

  This suggests that, the enhancement of stability of metastable state
is strongly correlated with the potential fluctuations, when the
Brownian particle ''\emph{sees}'' the barrier of the metastable
state~\cite{man96,dub04,agu01}. When the average time to cross the
barrier, that is the average lifetime of the metastable state, is
approximately equal to the correlation time of the fluctuations of
the potential barrier, a resonance-like phenomenon occurs. In other
words, this new effect can be considered as a NES effect in the
frequency domain. It is worthwhile to note that the new nonmonotonic
behaviour shown in Fig.~\ref{fig3} is in good agreement with
experimental results observed in a periodically driven Josephson
junction (JJ)~\cite{sun06}. In this very recent paper the authors
experimentally observe the coexistence of RA and NES phenomena.
Specifically they found (see Fig.$3$ of the paper~\cite{sun06}) that
the maximum increases with decreasing bias current and at the same
time the position of the maximum is shifted towards lower values of
$\omega$. A decrease of the bias current causes (see next section on
transient dynamics of a JJ) a decrease of the slope of the potential
profile, which corresponds to a decreasing parameter $k$ in our
model (Eq.~(\ref{piecewise})). Therefore, the average lifetime
maximum $\tau_{-_{max}}$ increases and as a consequence the time
required to escape from the metastable fixed configuration
$\tau_{up}(D)$ increases too. Consequently, the corresponding rate
of the barrier fluctuations $\omega_{max}\simeq 1/\tau_{up}(D)$
decreases, as observed experimentally. Of course a more detailed
analysis of the JJ system as a function of the temperature, that is
the noise intensity, should add more interesting results.

Finally we note that in the frequency range $\omega \in (10^{-5}
\div 10^{-3})$, for fixed values of the mean flipping rate, an
overlap occurs in the curves for different values of the noise
intensity. A nonmonotonic behavior of $\tau_{-}(x_{in} = 0)$ as a
function of the noise intensity is observed, as we expect in the
transient dynamics of metastable states~\cite{man96,dub04,agu01}.

\section{Transient dynamics in a Josephson junction}
The investigation of thermal fluctuations and nonlinear properties
of Josephson junctions (JJs) is very important owing to their broad
applications in logic devices. Superconducting devices, in fact, are
natural qubit candidates for quantum computing because they exhibit
robust, macroscopic quantum behavior~\cite{mak01}. Recently, a lot
of attention was devoted to Josephson logic devices with high
damping because of their high-speed switching~\cite{pan04,ort03}.
The rapid single flux quantum logic (RSFQ), for example, is a
superconductive digital technique in which the data are represented
by the presence or absence of a flux quantum $\Phi_0 = h/2e$, in a
cell which comprises Josephson junctions. The short voltage pulse
corresponds to a single flux quantum moving across a Josephson
junction, that is a $2 \pi$ phase flip. This short pulse is the unit
of information. However the operating temperatures of the high-Tc
superconductors lead to higher noise levels by increasing the
probability of thermally-induced switching errors. Moreover during
the propagation within the Josephson transmission line fluxon
accumulates a time jitter. These noise-induced errors are one of the
main constraints to obtain higher clock frequencies in RSFQ
microprocessors~\cite{ort03}.

In this section, after a short introduction with the basic formulas
of the Josephson devices, the model used to study the dynamics of a
short overdapmed Josephsonn junction is described. The interplay of
the noise-induced phenomena RA and NES on the temporal
characteristics of the Josephson devices is discussed. The role
played by these noise-induced effects, in the accumulation of timing
errors in RSFQ logic devices, is analyzed.

\begin{figure}[htp]
\centerline{\epsfxsize=2.5in\epsfbox{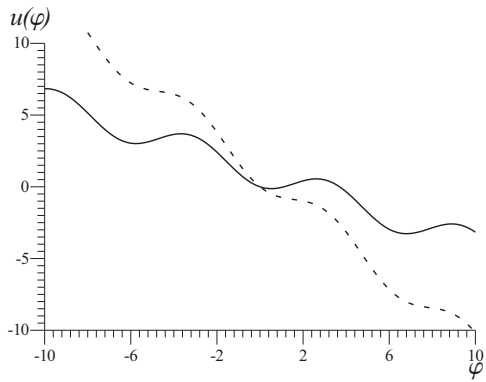}} \caption{The
potential profile $u(\varphi)=1-\cos \varphi-i\varphi$, for values
of the current, namely $i=0.5$ (solid line) and $i=1.2$ (dashed
line).} \label{fig4}
\end{figure}

The Josephson tunneling junction is made up of two superconductors,
separated from each other by a thin layer of oxide. Starting from
Schr\"{o}dinger equation and the two-state approximation
model~\cite{bar82}, it is straightforward to obtain the Josephson
equation

\begin{equation}
\frac{d\varphi(t)}{dt} = \frac{2 e V(t)}{\hbar},
\label{J eq.}
\end{equation}
where $\varphi$ is the phase difference  between the wave function
for the left and right superconductors, $V(t)$ is the potential
difference across the junction, $e$ is the electron charge, and
$\hbar = h/2\pi$ is the Planck's constant. A small junction can be
modelled by a resistance R in parallel with a capacitance C, across
which is connected a bias generator and a phase-dependent current
generator, $I sin \varphi$, representing the Josephson supercurrent
due to the Cooper pairs tunnelling through the junction. Since the
junction operates at a temperature above absolute zero, there will
be a white Gaussian noise current superimposed on the bias current.
Therefore the dynamics of a short overdamped JJ, widely used in
logic elements with high-speed switching and corresponding to a
negligible capacitance C, is obtained from Eq.~(\ref{J eq.}) and
from the current continuity equation of the equivalent circuit of
the Josephson junction. The resulting equation is the following
Langevin equation

\begin{equation}
\omega _c^{-1}{\frac{d\varphi (t)}{dt}}=-{\frac{du(\varphi )}{
d\varphi }}-i_F(t),
\label{Lang eq.JJ}
\end{equation}
valid for $\beta \ll 1$, with $\beta =2eI_cR^2C/{\hbar }$ the
McCamber--Stewart parameter, $I_c$ the critical current, and
$\displaystyle{i_F(t)={\frac{I_F }{I_c}}}$, with $I_F$ the random
component of the current. Here

\begin{equation}
u(\varphi, t)=1-cos \varphi -i(t) \varphi, \,\,\,\, \mathrm{with}
\;\;\; i(t) = i_0+f(t), \label{potJJ}
\end{equation}
is the dimensionless potential profile (see Fig.~\ref{fig4}),
$\varphi $ is the difference in the phases of the order parameter on
opposite sides of the junction, $f(t) = A\sin(\omega t)$ is the
driving signal, $\displaystyle{i={\frac I{I_c}}}$, $\displaystyle
\omega_c={\frac{2eR_NI_c}\hbar }$ is the characteristic frequency of
the JJ, and $R_N$ is the normal state resistance (see
Ref.~\cite{bar82}). When only thermal fluctuations are taken into
account \cite{bar82}, the random current may be represented by the
white Gaussian noise: $\left<i_F(t)\right>=0$, \quad $\displaystyle
{\left <i_F(t)i_F(t+\tau )\right>}$= $\displaystyle
{\frac{2D}{{\omega _c}}}\delta (\tau )$, where $\displaystyle D =
{\frac{2ekT}{{\hbar I_c}}}={\frac{I_T}{{I_c}}}$ is the dimensionless
intensity of fluctuations, $T$ is the temperature and $k$ is the
Boltzmann constant.
The equation of motion Eq.~(\ref{Lang eq.JJ}) describes the
overdamped motion of a Brownian particle moving in a washboard
potential (see Fig.~\ref{fig4}). A junction initially trapped in a
zero-voltage state, with the particle localized in one of the
potential wells, can escape out of the potential well by thermal
fluctuations. The phase difference $\varphi$ fluctuates around the
equilibrium positions, minima of the potential $u(\varphi )$, and
randomly performs jumps of $2\pi$ across the potential barrier
towards a neighbor potential minimum. The resulting time phase
variation produces a nonzero voltage across the junction with marked
spikes. For a bias current less than the critical current $I_c$,
these metastable states correspond to "superconductive" states of
the JJ. The mean time between two sequential jumps is the life time
of the superconductive metastable state \cite{mal96}. For an
external current greater than $I_c$, the JJ junction switches from
the superconductive state to the resistive one and the phase
difference slides down in the potential profile, which now has not
equilibrium steady states. A Josephson voltage output will be
generated in a later time. Such a time is the switching time, which
is a random quantity. In the presence of thermal noise a Josephson
voltage appears even if the current is less than the critical one
($i < 1$), therefore we can identify the lifetime of the metastable
states with the mean switching time \cite{pan04,mal96}. For the
description of our system, i. e. a single overdamped JJ with noise,
we will use the Fokker-Planck equation for the probability density
$W(\varphi ,t)$, which corresponds to the Langevin equation
(\ref{Lang eq.JJ})

\begin{eqnarray}
{\partial W(\varphi,t)\over\partial t}= -{\partial
G(\varphi,t)\over\partial\varphi}= {\omega_c}{\partial\over\partial
\varphi}\left\{ {du(\varphi)\over d\varphi}W(\varphi,t)+ D {\partial
W(\varphi,t)\over\partial \varphi}\right\}. \label{FP eq.}
\end{eqnarray}
The initial and boundary conditions of the probability density and
of the probability current for the potential profile (\ref{potJJ})
are as follows: $W(\varphi,0)=\delta(\varphi-\varphi_0)$,
$W(+\infty,t)=0,$ $G(-\infty,t)=0$. Let, initially, the JJ is biased
by the current smaller than the critical one, that is $i_0<1$, and
the junction is in the superconductive state. The current pulse
$f(t)$, such that $i(t)=i_0+f(t)>1$, switches the junction into the
resistive state. An output voltage pulse will appear after a random
switching time. We will calculate the mean value and the standard
deviation of this quantity for two different periodic driving
signals: (i) a dichotomous signal, and (ii) a sinusoidal one. We
will consider different values of the bias current $i_o$ and of
signal amplitude $A$. Depending on the values of $i_o$ and $A$, as
well as values of signal frequency and noise intensity, two
noise-induced effects may be observed, namely the resonant
activation (RA) and the noise enhanced stability (NES). Specifically
the RA effect was theoretically predicted in Ref.~\cite{doe92} and
experimentally observed in a tunnel diode~\cite{man00} and in
underdamped Josephson tunnel junctions~\cite{yu03,sun06}, and the
NES effect was theoretically predicted in~\cite{hir82,agu01,mal96}
and experimentally observed in a tunnel diode~\cite{man96} and in an
underdamped Josephson junction~\cite{sun06}. The RA and NES effects,
however, have different role on the behavior of the temporal
characteristics of the Josephson junction. They occur because of the
presence of metastable states, in the periodic potential profile of
the Josephson tunnel junction, and the thermal noise. Specifically,
the RA phenomenon minimizes the switching time and therefore also
the timing errors in RSFQ logic devices, while the NES phenomenon
increases the mean switching time producing a negative effect
\cite{pan04}.

\subsection{Temporal characteristics}

Now we investigate the following temporal characteristics: the mean
switching time (MST) and its standard deviation (SD) of the
Josephson junction described by Eq.~(\ref{Lang eq.JJ}). These
quantities may be introduced as characteristic scales of the
evolution of the probability
$P(t)=\int\limits_{\varphi_1}^{\varphi_2} W(\varphi,t)d \varphi$, to
find the phase within one period of the potential profile of Eq.
(\ref{potJJ}). We choose therefore $\varphi_2=\pi$, $\varphi_1=-\pi$
and we put the initial distribution on the bottom of a potential
well: $\varphi_0=\arcsin(i_0)$. A widely used definition of such
characteristic time scales is the integral relaxation time (see the
paper by Malakhov and Pankratov in Ref~\cite{bie93}). Let us
summarize shortly the results obtained in the case of dichotomous
driving, $f(t)=A {\rm sign}(\sin(\omega t))$. Both MST and its SD do
not depend on the driving frequency below a certain cut-off
frequency (approximately $0.2 \omega_c$), above which the
characteristics degrade. In the frequency range from $0$ to $0.2
\omega_c$, therefore, we can describe the effect of dichotomous
driving by time characteristics in a constant potential. The exact
analytical expression of the first two moments of the switching time
are~\cite{pan04}

\begin{equation}
\tau_c(\varphi_0)=\frac{1}{D \omega _c} \left[
\int\limits_{\varphi_0}^{\varphi_2}e^{\frac{u(x)}{D}}
\int\limits_{\varphi_1}^xe^{-\frac{u(\varphi)}{D} } d\varphi dx+
\int\limits_{\varphi_1}^{\varphi_2} e^{-\frac{u(\varphi)}{D}}
d\varphi \int\limits_{\varphi_2}^\infty e^{\frac{u(\varphi)}{D}
}d\varphi \right],
\label{tauc}
\end{equation}
and

\begin{equation}
\tau_{2c}(\varphi_0) = \tau_c^2(\varphi_0)-
\int_{\varphi_0}^{\varphi_2}e^{-\frac{u(x)}{D}} H(x)dx -
H(\varphi_0)\int_{\varphi_1}^{\varphi_0} e^{-\frac{u(x)}{D}}dx,
\label{tau2c}
\end{equation}
where $ H(x)= \frac{2}{(D\omega _c)^2}
\int\limits_{x}^{\infty}e^{u(v)/D }
\int\limits_{v}^{\varphi_2}e^{-u(y)/D }
\int\limits_{y}^{\infty}e^{u(z)/D} dzdydv$. The asymptotic
expressions of the MST and its standard deviation (SD), obtained in
the small noise limit ($D \ll 1$), agree very well with computer
simulations up to $D = 0.05$~\cite{pan04}. Therefore, not only low
temperature devices ($D\leq 0.001$), but also high temperature
devices may be described by these expressions. If the noise
intensity is rather large, the phenomenon of NES may be observed in
our system: the MST increases with the noise intensity. Here we note
that it is very important to consider this effect in the design of
large arrays of RSFQ elements, operating at high frequencies. To
neglect this noise-induced effect in such nonlinear devices it may
lead to malfunctions due to the accumulation of errors.

\begin{figure}[htp]
\centerline{\epsfxsize=2.5in\epsfbox{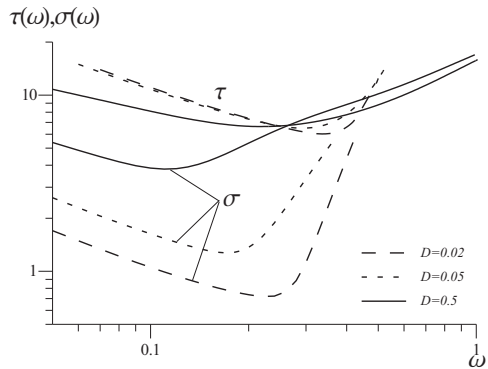}} \caption{The MST
and its SD vs frequency for $f(t)=A\sin(\omega t)$ (computer
simulations) for three values of the noise intensity. Namely:
Long-dashed line - $D=0.02$, short-dashed line - $D=0.05$, solid
line - $D=0.5$. The value of the bias current is $i_0 =0.5$, and
 the total current is $i =1.5$.
 \label{fig5}}
\end{figure}

Now let us consider the case of sinusoidal driving. The
corresponding time characteristics may be derived using the modified
adiabatic approximation \cite{pan00,pan04}

\begin{equation}
P(\varphi_0,t)=\exp\left\{-\int_{0}^t
\frac{1}{\tau_c(\varphi_0,t')}dt'\right\}, \label{met}
\end{equation}
with $\tau_c(\varphi_0,t')$ given by Eq.~(\ref{tauc}), after
inserting in this equation the time dependent potential profile
$u(\varphi,t)$ of Eq.~(\ref{potJJ}). Using the relation $\tau =
\int_{0}^{+\infty} P(\varphi_0,t) dt$ we calculate the MST. We focus
now on the current value $i=1.5$, because $i=1.2$ is too small for
high frequency applications. In Fig.~\ref{fig5} the MST and its SD
as a function of the driving frequency, for three values of the
noise intensity ($D = 0.02, 0.05, 0.5$), for a bias current $i_0 =
0.5$, and $A=1$ are shown. We note that, because
$\varphi_0=\arcsin(i_0)$ depends on $i_0$, the switching time is
larger for smaller $i_0$. However, great bias current values $i_0$,
in the absence of driving, give rise to the reduction of the mean
life time of superconductive state, i.e. to increasing storage
errors (Eq.~(\ref{tauc})). Therefore, there must be an optimal value
of bias current $i_0$, giving minimal switching time and acceptably
small storage errors. We observe the phenomenon of resonant
activation: MST has a minimum as a function of driving frequency.

\begin{figure}[htp]
\centerline{\epsfxsize=2.5in\epsfbox{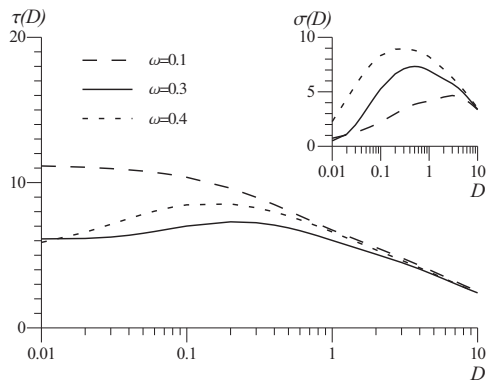}} \caption{The MST vs
noise intensity for $f(t)=A\sin(\omega t)$ and for three values of
the driving frequency. Namely: $\omega = 0.1$ (long-dashed line),
$\omega = 0.3$ (short-dashed line), $\omega = 0.4$ (solid line).
Inset: The standard deviation (SD) vs noise intensity for the same
values of driving frequency $\omega$.
\label{fig6}}
\end{figure}

The approximation (\ref{met}) works rather well below 0.1
$\omega_c$, that is enough for practical applications (see the inset
of Fig.~3 in ref.~\cite{pan04}). It is interesting to see that near
the minimum the MST has a very weak dependence on the noise
intensity (as it is clearly shown in the $\tau$ behavior of
Fig.~\ref{fig5} for three values of the noise intensity), i. e. in
this signal frequency range the noise is effectively suppressed.
This noise suppression is due to the resonant activation phenomenon:
a minimum appears in the MST and SD, when the escape process is
strongly correlated with the potential profile oscillations. A noise
suppression effect, but due to the noise, is reported in
Ref.~\cite{vil01}. We observe also the NES phenomenon. There is a
frequency range in Fig.~\ref{fig5}, around $(0.2 \div 0.48)
\omega_c$ for $i_0=0.5$, where the switching time increases with the
noise intensity. To see in more detail this effect we report in
Fig.~\ref{fig6} the MST $\tau(D)$ and its SD $\sigma(D)$ vs the
noise intensity $D$, for three fixed values of the driving
frequency, namely: $\omega = 0.1, 0.3, 0.4$. Both quantities have
nonmonotonic behaviour and the great values of $\sigma(D)$ near the
maximum of $\tau(D)$ confirm that the only information on the MST is
not sufficient to fully unravel the statistical properties of the
NES effect~\cite{fia05}. A detailed analysis of the PDF of the
lifetime during the transient dynamics is required. This is subject
of a forthcoming paper. Simulations for different bias current
values~\cite{pan04} show that the NES effect increases for smaller
$i_0$ because the potential barrier disappears for a short time
interval within the driving period $T = 2\pi/\omega$ and the
potential is more flat~\cite{man96}. The noise, therefore, has more
chances to prevent the phase to move down and the switching process
is delayed. This effect may be avoided, if the operating frequency
does not  exceed $0.2~\omega_c$. Besides the SD and MST (see
Fig.~\ref{fig5}) have their minima in a short range of values of
$\omega$~\cite{pan04}. Close location of minima of MST and its SD
means that optimization of RSFQ circuit for fast operation will
simultaneously lead to minimization of timing errors in the circuit.

\section{Dynamics of a FHN stochastic model}

\subsection{Suppression of noise and noise-enhanced stability effect}

{\it{Case I}.} Let us fix the value of the noise intensity and
analyze The analysis of the stochastic properties of neural systems
is of particular importance since it plays an important role in
signal transmission~\cite{panka05}, \cite{lee98}-\cite{fit61}.
Biologically realistic models of the nerve cells, such as
widely-known Hodgkin-Huxley (HH) system~\cite{lee98}, are so complex
that they provide little intuitive insight into the neuron dynamics
that they simulate. The FitzHugh-Nagumo (FHN) model, however, which
is one of the simplified modifications of HH, is more preferable for
investigation~\cite{fit61}. Nevertheless many effects observed in
neural cells are qualitatively contained in FHN model. Because of
this the FHN model has got wide dissemination in the last few years.
There has been a lot of papers where the influence of noise on the
encoding sensitivity of a neuron in the framework of FHN model has
been analyzed. A broad spectrum of noise-induced dynamical effects,
which produce ordered periodicity in the output of the FHN system,
has been discovered. Among these effects we cite the coherence
resonance~\cite{pik97} and the stochastic resonance
(SR)~\cite{lon98}. All these investigations deal with neuron
dynamics with subthreshold signals, and with an enhancement of a
weak signal through the noise. The presence of noise in the case of
a strong periodic forcing, however, has a detrimental effect on the
encoding process~\cite{bul96}-\cite{sto01}. For suprathreshold
signals the noise always lowers the information transmission, and
the SR effects disappear~\cite{bul96,lev96}. However, as it was
shown in recent papers of Stocks~\cite{sto01}, this is only true for
a single element threshold system. In neuronal arrays the noise can
significantly enhance the information transmission when the signal
is predominantly suprathreshold. It is the effect of suprathreshold
stochastic resonance.

Here we analyze the effect of noise in a single neuron subjected to
a strong periodic forcing. We investigate therefore, the influence
of noise on the appearance time of a first spike, or the mean
response time, in the output of FHN model with periodical driving in
suprathreshold regime. As it was mentioned before, the role of noise
for a strong driving is negative. In this case noise suppresses the
response of a neuron, that leads to delay of transmission of an
external information. But we show that, this negative influence of
noise on the spike generation can be significantly minimized.

We analyze the dependencies of the mean response time (MRT) on both
driving frequency and noise intensity. We find that, MRT plotted as
a function of the driving frequency shows a resonant activation-like
phenomenon. The noise enhanced stability (NES) effect is also
observed here. It is shown that MRT can be increased due to the
effect of fluctuations. We note that NES has nothing to do with the
typical SR, where the maximum of signal to noise ratio as a function
of noise intensity is observed. There are many differences between
these effects concerning the neuron dynamics. First of all the SR is
related to the output of the neuron in stationary dynamical regime
and concerns the signal-to-noise ratio, while the NES describes the
transient dynamical regime of a neuron and concerns the mean
response time. In addition there is difference in the nature of the
response: We investigate the case of a strong driving, where the SR
effects disappear.

\subsection{Deterministic dynamical regime}

The dynamic equations of the FitzHugh-Nagumo model with additive
periodic forcing are
\begin{eqnarray}
\begin{array}{lll}
\dot{x} & = & x - x^3/3 - y +A\sin(\omega t)\\
\dot{y} & =& \epsilon (x+I),
\end{array}
\label{1e}
\end{eqnarray}
where $x$ is the voltage, $y$ is the recovery variable, and
$\epsilon$ is a fixed small parameter ($\epsilon =0.05$). In the
absence of both external driving and noise, there is only one steady
state of the system~(\ref{1e}), that is $x_0=-I;~ y_0=-I+I^3/3$. The
choice of the constant $I$, therefore, fully specifies the location
of equilibrium state in the phase space $(x,y)$. Here we consider $I
= 1.1$.

In our simulations we assume that the initial conditions for each
realization are the same, that is the system is in its stable
equilibrium point (the rest state) $(x_0,y_0)$ at the initial time
$t_0$. We would like to note, here, that even if we consider
sinusoidal driving, we investigate the time of appearance of the
first spike only. We are interested in the capability of our system
to detect an external input. This means to minimize the detection
time and to get a neuron response that would be more robust to the
noise action. After generation of the first spike, that is after
approaching of the boundary $x=0$, we break the realization off, and
start a new one with the same initial conditions $(x_0,y_0)$.

\begin{figure}[h]
\begin{center}
\includegraphics[width=12cm,height=6cm]{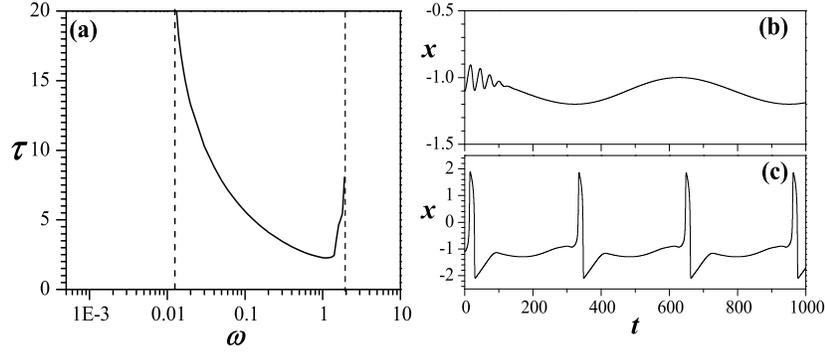}
\end{center}
\vskip -1.cm \caption{ {\small (a) The response time dependence
versus the frequency of periodic driving for the deterministic case,
$A=0.5$. Examples of output trajectory for two values of driving
frequency invoking two different kinds of oscillations: (b)
subthreshold for $\omega = 0.01$, and (c) suprathreshold for $\omega
= 0.02$. }} \label{fig7}
\end{figure}

For our system the threshold value of the driving amplitude required
for spike generation is $A_{th} \sim 0.05$. In our simulations we
choose $A=0.5$. Thus, the frequency range where the signal of such
amplitude is suprathreshold is: $\Omega :\omega \in (0.013 \div
1.9)$~\cite{panka05}. Inside this region $\Omega$ the response time
of a neuron has a minimum as a function of the driving frequency.
System does not respond outside the range $\Omega$. Here, a
subthreshold oscillation occurs (see Fig.~\ref{fig7}(b)). In
Fig.~\ref{fig7}(c) the time series of the output voltage $x$ for a
suprathreshold signal is shown.

\subsection{Suprathreshold stochastic regime}

Actually, there are many factors that make the environment noisy in
the neuron dynamics. Among them we cite the fluctuating opening and
closure of the ion channels within the cell membrane, the noisy
presynaptic currents, and others (see, for example,
Ref.~\cite{pik97}). We consider two different cases in which the
noise is added to the first or the second equation of the
system~(\ref{1e}):

{\underline{Case $I$}} The variable that corresponds to the membrane
potential is subjected to fluctuations~\cite{lon98,sto01}. In this
case, the first equation of the system~(\ref{1e}) becomes the
following stochastic differential equation
\begin{equation}
\dot{x}=x-x^3/3+A\sin(\omega t)-y+\xi(t); \label{1ea}
\end{equation}

{\underline{Case $II$} } The recovery variable associated with the
refractory properties of a neuron is noisy~\cite{pik97}. Here, the
second equation of the system~(\ref{1e}) becomes
\begin{eqnarray}
\displaystyle{\dot{y}=\epsilon (x+I)+\xi(t)}, \label{1eb}
\end{eqnarray}

\noindent In Eqs.~(\ref{1ea}) and (\ref{1eb}), $\xi(t)$ is a
Gaussian white noise with zero mean and correlation function
$\left<\xi(t)\xi(t+\tau)\right>=D\delta(\tau)$. For numerical
simulations we use the modified midpoint method and the noise
generator routine reported in Ref~\cite{pre93}.

\begin{figure}[h]
\centerline{\epsfxsize=4.5in\epsfbox{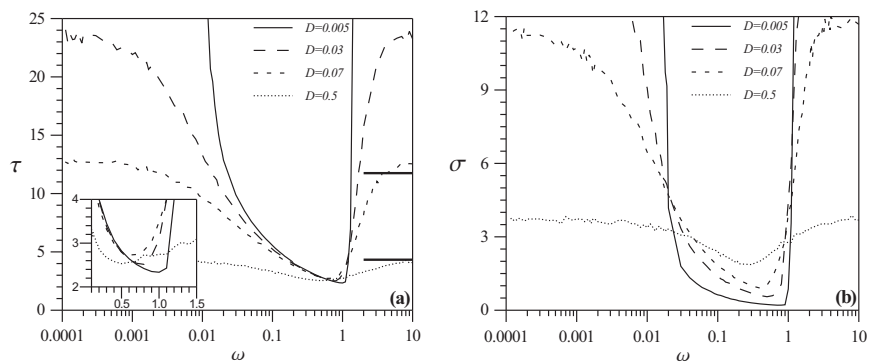}} \caption{ {\small
(a) The mean response time dependence versus the frequency of
periodic driving for case $I$, for four values of the noise
intensity, namely: $D = 0.005,0.03,0.07,0.5$. The right solid lines
give the theoretical values of $\tau$ for fixed bistable potential.
Inset: frequency range where the noise enhanced stability effect is
observed. (b) The standard deviation of the response time dependence
vs frequency of periodic driving for case $I$, for the same values
of $D$. }} \label{fig8}
\end{figure}

The mean response time (MRT) of our neuronal system is obtained as
the mean first passage time at the boundary $x=0$: $\tau=<T> =
\frac{1}{N}\sum_{i=1}^{N}T_i,$ where $T_i$ is the response time for
$i$-th realization. To obtain smooth average for all the noise
values investigated, we need different number of realizations $N$ in
above considered cases. Namely, $N=5000$ in case $I$, and $N=15000$
in case $II$, specifically when the noise intensities are comparable
with the value of the parameter $\epsilon = 0.05$. It is worth
noting here that parameter $T_i$ characterizes the delay of the
systems' response, and has a non-zero value even in the
deterministic case, because of the non-instantaneous neuronal
response. In our investigation we consider a strong driving, so the
noise increases the time of appearance of the first spike and leads
to an additional delay of the signal detection.

the MRT dependence on the driving frequency. In the small noise
limit $D\rightarrow 0$, a typical behavior (see Fig.~\ref{fig8})
with perpendicular walls disposed at the frequencies corresponding
to the boundaries of the region $\Omega$ was found. By increasing
the noise intensity, these walls go down. We observe a resonant
activation-like phenomenon: The MRT exhibits a minimum as a function
of the driving frequency, which is almost independent of the noise
intensity.

In the same figure (b) the standard deviation (SD) of the response
time versus the frequency of the periodic driving shows a minimum.
Therefore, the noise has minimal effect in the same range of driving
frequencies for MRT and its SD. In a narrow frequency range ($\omega
\in (0.6\div 1.3)$) (see Fig.~\ref{fig8}), we found a nonmonotonic
behavior of the MRT as a function of the noise intensity. Here the
noise enhanced stability effect is observed (see the inset of
Fig.~\ref{fig8}a). Out of this range the MRT monotonically decreases
with increasing noise intensity. For larger noise intensities the
MRT dependence on driving frequency takes a constant-like behavior
in the range of the investigated frequency values ($\omega \in
[10^{-4} \div 10]$). Here, the dynamics of the system is mainly
controlled by the noise, and the frequency of periodic driving does
not affect significantly the neuron response dynamics. By numerical
simulations of our system we find that, for large noise intensities,
the MRT coincides with that calculated by standard technique for a
Brownian particle moving in a bistable fixed potential~\cite{gar04}

\begin{equation}
\tau = 2/D
\int_{x_0}^{0}e^{\varphi(x)/D}\int_{\infty}^{x}e^{-\varphi(y)/D}dy
dx.
\end{equation}
The theoretical values reported in Fig.~\ref{fig8}(a) agree with the
limiting values of $T$ for $\omega\rightarrow 0$ and
$\omega\rightarrow \infty$. For $\epsilon \ll 1$ in fact, $x$ is a
fast variable and $y$ is a slow variable, so $\dot{y}\simeq 0$ and
this case can be recast as an escape problem from a one-dimensional
double well in both limiting cases. In fact when $\omega \rightarrow
0$ we have a fixed bistable potential, and for $\omega \rightarrow
\infty$ we have an average fluctuating potential, which coincides
with the fixed one. This is well visible in Fig.~\ref{fig8}(a) for
$D = 0.07$. For $D = 0.5$ the MRT tends to be almost independent on
the parameter $\omega$.

\begin{figure}[h]
\centerline{\epsfxsize=2.5in\epsfbox{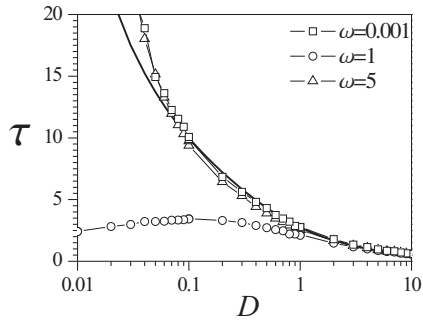}}\ \caption{ {\small
The mean response time dependence versus the noise intensity for
case $I$, for three different values of driving frequency: $\omega
=0.001$, $\omega =1$ and $\omega =5$. Solid line gives the
theoretical values of $\tau$ for fixed bistable potential.}}
\label{fig9}
\end{figure}
In Fig.~\ref{fig9} the MRT versus the noise intensity, for three
values of the driving frequency, is shown. We see the nonmonotonic
behavior for $\omega = 1$, which is a signature of the NES effect.
It is interesting to note that even in this system, whose global
dynamics cannot be described as the motion of a Brownian particle in
a potential profile (because of the coupling between the two
stochastic differential equations describing our system) a phase
transition-like phenomenon, with respect to the driving frequency
parameter $\omega$, occurs. In fact we have nonmonotonic and
monotonic behavior depending on the value of $\omega$. We expect
similar behavior, if we fix the driving frequency and we change the
value of the amplitude $A$ of the driving force~\cite{man96,dub04}.

{\it{Case II}.} In this case, for noise intensity values greater
than $\epsilon =0.05$, the recover variable can be approximated by a
Wiener process ($\dot{y}\approx \xi(t)$). This process acts now as a
noise source in the same double well potential according the
following stochastic differential equation

\begin{equation}
\dot{x}=x-x^3/3+A\sin(\omega t) + W(t) \, ,
\end{equation}
where $W(t)$ is the Wiener process with the usual statistical
properties: $\langle W(t) \rangle = 0$, and $\langle W^2 (t)\rangle
= t$.
\begin{figure}[h]
\centerline{\epsfxsize=4.5in\epsfbox{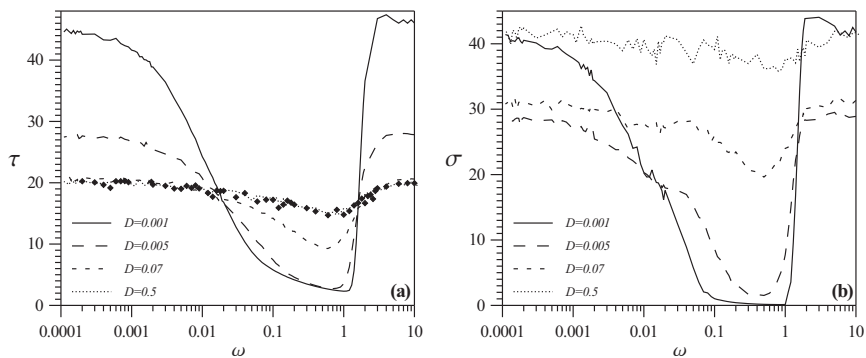}} \caption{ {\small
(a) The mean response time dependence vs the frequency of the
periodic driving for case $II$, for four values of the noise
intensity, namely:  $D = 0.001,0.005,0.07,0.5$. The curve with
diamonds gives the values of $\tau$ for fixed bistable potential,
when the noise source is a Wiener process. (b) The standard
deviation of the response time dependence vs frequency of periodic
driving for case $II$. \vspace{-0.3cm}}} \label{fig10}
\end{figure}

\noindent In Fig.~\ref{fig10}(a) the curve with diamonds shows the
results of this approximation for $D=0.5$. We found again a resonant
activation-like phenomenon (see Fig.~\ref{fig10}), which is
independent of the noise intensity, as in case $I$, until $D$
reaches the value of parameter $\epsilon$. The minimum tends to
disappear for greater noise intensities. Here a certain frequency
range ($\omega \in (0.019 \div 1.6)$), larger than in previous case,
exists where an increasing noise intensity leads to a monotonic
growth of the MRT. Out of this range the MRT monotonically decreases
with increasing noise intensity, as in case $I$. In
Fig.~\ref{fig10}(b) the standard deviation of the response time
versus the driving frequency is shown. Also in this case $II$, the
SD shows a minimum in the same frequency range of that found for
MRT. We have found, therefore, a parameter region where there is
minimization of the MRT and its SD, that is ''\emph{suppression of
noise}''.

We also observe that the saturation level reached in each case is
different. Particularly in case $II$ it is greater than in case $I$,
because the MRT is calculated with respect to the membrane voltage
$x$ and with different noise sources. Therefore, in phase space the
variable $x$ reaches, in case $I$, in a minor average time the
boundary $x=0$, according to Eq.~(\ref{1ea}). While in case $II$ the
variation of $x$ depends on the dynamics of the $y$ coordinate and
takes much more time to reach the same boundary.

\section{A stochastic model for cancer growth dynamics}

In this last section we shortly summarize some of the main results
obtained with a stochastic model for cancer growth dynamics (see
Ref.~\cite{ale06} for more details). Most of tumoral cells bear
antigens which are recognized as strange by the immune system. A
response against these antigens may be mediated either by immune
cells such as T-lymphocytes or other cells like macrophages. The
process of damage to tumor proceeds via infiltration of the latter
by the specialized cells, which subsequently develop a cytotoxic
activity against the cancer cell-population. The series of cytotoxic
reactions between the cytotoxic cells and the tumor tissue have been
documented to be well approximated by a saturating, enzymatic-like
process whose time evolution equations are similar to the standard
Michaelis-Menten kinetics~\cite{gar78,och04}. The T-helper
lymphocytes and macrophages, can secrete cytokines in response to
stimuli. The functions that cytokines induce can both "\emph{turn
on}" and "\emph{turn off}" particular immune
responses~\cite{sic04,ell05}. This "\emph{on-off}" modulating
regulatory role of the cytokines is here modelled through a
dichotomous random variation of the parameter $\beta$, which is
responsibile for regulatory inhibition of the population growth, by
taking into account the natural random fluctuations always present
in biological complex systems.

The dynamical equation of this biological system is

 \begin{equation}
\dot{x} = -\frac{dU^\pm(x)}{dx} + \xi(t),
\label{eq:langevin}
\end{equation}
where $\xi(t)$ is a Gaussian process with $\langle \xi(t)\rangle =
0$, $\langle \xi(t)\xi(t')\rangle= D \delta(t-t')$, and

\begin{equation}
U^\pm(x)=-\frac{x^2}{2}+\frac{\theta x^3}{3}+(\beta_o \pm \Delta)( x
- \ln(x+1)),
\label{MM stoc pot}
\end{equation}
is the stochastic double well Michaelis-Menten potential with one
the minima at $x=0$. Here $x(t)$ is the concentration of the cancer
cells. The process $\beta = (\beta_o \pm \Delta)$ can change the
relative stability of the metastable state of the potential
profile~\cite{och04}. We note that the RA and NES phenomena act
counter to each other in the cancer growth dynamics: the NES effect
increases in an unavoidable way the average lifetime of the
metastable state (associated to a fixed-size tumor state), while the
RA phenomenon minimizes this lifetime. Therefore it is crucial to
find the optimal range of parameters in which the positive role of
resonant activation phenomenon, with respect to the cancer
extinction, prevails over the negative role of NES, which enhances
the stability of the tumoral state. These are just the main results
of the paper~\cite{ale06}, that is both NES and RA phenomena are
revealed in a biological system with a metastable state, with a
co-occurrence region of these effects. In this coexistence region
the NES effect, which enhances the stability of the tumoral state,
becomes strongly reduced by the RA mechanism, which enhances the
cancer extinction. In other words, an asymptotic regression to the
zero tumor size may be induced by controlling the modulating
stochastic activity of the cytokines on the immune system.

\section{Conclusions}

Natural systems are open to the environment. Consequently, in
general, stationary states are not equilibrium states, but are
strongly influenced by dynamics, which adds further challenge to the
microscopic understanding of metastability. The investigation of two
noise-induced effects in far from equilibrium systems, namely the RA
and NES phenomena, has revealed interesting peculiarities of the
dynamics of these systems. Specifically the knowledge of the
parameter regions where the RA and NES can be revealed allows:

\begin{itemize}
\item to optimize and to suppress "timing" errors in practical RSFQ
devices, and therefore to significantly increase working frequencies
of RSFQ circuits;

\item to optimize the operating range of a neuron, and therefore
 to realize high rate signal transmission with the suppression of
noise;

\item to maximize or minimize the extinction time in cancer growth
 population dynamics.

\end{itemize}

\section{Acknowledgments} \vskip-0.2cm
This work was supported by MIUR, INFM-CNR and CNISM, Russian
Foundation for Basic Research (Projects No. 05-01-00509 and No.
05-02-19815), and ESF(European Science Foundation) STOCHDYN network.
E.V.P. also acknowledges the support of the Dynasty Foundation.

%\newpage


\begin{thebibliography}{99}

\bibitem{gun83}
J.D. Gunton, M. Droz, \emph{Introduction to the Theory of Metastable
and Unstable States}, Springer, Berlin, 1983.

\bibitem {leg84}
A.~J.~Leggett, \emph{Phys. Rev. Lett.} \textbf{53}, 1096 (1984); M.
Muthukumar, \textit{ibid.} \textbf{86}, 3188 (2001).

\bibitem {man96}R. N. Mantegna and B. Spagnolo,
\emph{Phys. Rev. Lett.} \textbf{76}, 563 (1996); \emph{Int. J.
Bifurcation and Chaos} \textbf{4}, 783 (1998).

\bibitem{str99}
A. Strumia, N. Tetradis, \emph{JHEP} \textbf{11}, 023 (1999).

\bibitem{deb01}
P.G. Debenedetti, F.H. Stillinger, \emph{Nature} \textbf{410}, 267
(2001).

\bibitem{vic89}
R. H. Victora, \emph{Phys. Rev. Lett.} \textbf{63}, 457 (1989).

\bibitem{bai95}
 Y.W. Bai et al, \emph{Science} \textbf{269}, 192 (1995).

\bibitem{tre03}
O. A. Tretiakov, T. Gramespacher, and K. A. Matveev, \emph{Phys.
Rev. B} \textbf{67}, 073303 (2003).

\bibitem{gle05}
M. Gleiser, R.C. Howell, \emph{Phys. Rev. Lett.} \textbf{94}, 151601
(2005).

\bibitem{par05}
G. Parisi, \emph{Nature} \textbf{433}, 221  (2005); S. Kraut and C.
Grebogi,\emph{ Phys. Rev. Lett.} \textbf{93}, 250603  (2004); H.
Larralde and F. Leyvraz, \emph{ibid.} \textbf{94}, 160201  (2005);
G. B$\acute{a}$ez et al., \emph{ibid.} \textbf{90}, 135701 (2003).

\bibitem{doe92}
C. R. Doering and J. C. Gadoua, \emph{Phys. Rev. Lett.} \textbf{69},
2318 (1992).

\bibitem{bie93} M. Bier and R. D. Astumian,
\emph{Phys. Rev. Lett.} \textbf{71}, 1649 (1993); P. Pechukas and P.
H\"anggi, \textit{ibidem }\textbf{ 73}, 2772 (1994); J.
Iwaniszewski, \emph{Phys. Rev. E }\textbf{54}, 3173 (1996); M.
Bogu\~n\'a, J. M. Porra, J. Masoliver, and K. Lindenberg,
\textit{ibidem } \textbf{57}, 3990 (1998); M. Bier, I. Derenyi, M.
Kostur, D. Astumian, \textit{ibidem } \textbf{59}, 6422 (1999); A.
L. Pankratov and M. Salerno, \emph{Phys. Lett. A} {\bf 273}, 162
(2000); A. N. Malakhov and A. L. Pankratov, \emph{Adv. Chem. Phys.}
{\bf 121}, 357 (2002).

\bibitem{man00}
R. N. Mantegna and B. Spagnolo, \emph{Phys. Rev. Lett.} \textbf{84},
3025 (2000); \emph{J. Phys. IV (France)} \textbf{8}, 247 (1998).

\bibitem{pan00} A. L. Pankratov and M. Salerno, \emph{Phys.
Lett. A }{\bf 273}, 162 (2000).

\bibitem{dyb02}
B. Dybiec, E. Gudowska--Nowak,\emph{ Phys. Rev. E} \textbf{66},
026123 (2002).

\bibitem{yu03}
Y. Yu and S. Han, \emph{Phys. Rev. Lett.} {\bf 91}, 127003 (2003).

\bibitem{dub04}
A. A. Dubkov, N. V. Agudov and B. Spagnolo, \emph{Phys. Rev. E}
\textbf{69}, 061103 (2004).

\bibitem{pan04} A. L. Pankratov and B. Spagnolo,
\emph{Phys. Rev. Lett.} \textbf{93}, 177001 (2004).

\bibitem{panka05}
E. V. Pankratova, A. V. Polovinkin, and B. Spagnolo, \emph{Physics
Letters A} \textbf{344}, 43-50 (2005).

\bibitem{sun06}
G. Sun \emph{et al.}, \emph{Phys. Rev. E} \textbf{75}, 021107(4)
(2007).

\bibitem{ale06}
A. Fiasconaro and B. Spagnolo, A. Ochab-Marcinek and E.
Gudowska-Nowak, \emph{Phys. Rev. E } \textbf{74}, 041904(10) (2006);
A. Ochab-Marcinek, E. Gudowska-Nowak, A. Fiasconaro and B. Spagnolo,
\emph{Acta Physica Polonica B} \textbf{37} (5), 1651 (2006).

\bibitem {hir82}
J. E. Hirsch, B. A. Huberman, and D. J. Scalapino, \emph{Phys. Rev.
A} \textbf{25}, 519 (1982); I. Dayan, M. Gitterman, and G. H. Weiss,
\textit{ibidem } \textbf{46}, 757 (1992); R. Wackerbauer,
\emph{Phys. Rev. E} \textbf{59}, 2872 (1999); D. Dan, M. C. Mahato,
and A. M. Jayannavar, \textit{ibidem } \textbf{60}, 6421 (1999); A.
Mielke, \emph{Phys. Rev. Lett.} \textbf{84}, 818 (2000); C. Xie and
D. Mei, \emph{Chin. Phys. Lett.} \textbf{20}, 813 (2003).

\bibitem{agu01}
N. V. Agudov and B. Spagnolo, \emph{Phys. Rev. E} \textbf{64},
035102(R) (2001).

\bibitem{fia05}
A. Fiasconaro, B. Spagnolo and S. Boccaletti, \emph{Phys. Rev. E}
\textbf{72}, 061110(5) (2005); A. Fiasconaro, D. Valenti, B.
Spagnolo, \textit{Physica A} \textbf{325}, 136-143 (2003).

\bibitem{mal96} A. N. Malakhov and A.L. Pankratov,
\emph{Physica C} {\bf 269}, 46 (1996).

\bibitem{agu99}
N. V. Agudov and A. N. Malakhov, \emph{Phys. Rev. E} \textbf{60},
6333 (1999).

\bibitem{spa04} B. Spagnolo, A. A. Dubkov, and N. V. Agudov,
\emph{Eur. Phys. J. B} \textbf{40}, 273-281 (2004); B. Spagnolo, A.
A. Dubkov, N. V. Agudov, \emph{Acta Physica Polonica B} \textbf{35},
1419 (2004).

\bibitem{kra40}
H. A. Kramers, \emph{Physica} \textbf{7}, 284 (1940).

\bibitem{han90}
P. H\"{a}nggi, P. Talkner, and M. Borkovec, \emph{Rev. Mod. Phys.}
\textbf{62}, 251 (1990).

\bibitem{mal97}
A. N. Malakhov, \emph{Chaos} \textbf{7}, 488 (1997).

\bibitem{mak01} Y. Makhlin, G. Sch\"{o}n, and A. Shnirman,
\emph{Rev. Mod. Phys.} \textbf{73}, 357 (2001); Y. Yu \emph{et al.},
\emph{Science} \textbf{296}, 889 (2002).

\bibitem{ort03} T. Ortlepp, H. Toepfer and H. F. Uhlmann,
\emph{IEEE Trans. Appl. Supercond.} {\bf 13}, 515 (2003);
 V. Kapluneko, {\it Physica C} {\bf 372-376}, 119 (2002).

\bibitem{bar82} A. Barone and G. Paterno, {\it
Physics and Applications of the Josephson Effect}, Wiley, 1982.

\bibitem{vil01}
J. M. G. Vilar and J. M. Rub\'{i}, \emph{Phys. Rev. Lett.}
\textbf{86}, 950 (2001).

\bibitem{lee98}
S. Lee, A. Neiman, and S. Kim, \emph{Phys. Rev. E} \textbf{57}, 3292
(1998).

\bibitem{pik97}
A.S.~Pikovsky and J.~Kurths, \emph{Phys. Rev. Lett.} \textbf{78},
775 (1997); B.~Lindner and L.~Schimansky-Geier, \emph{Phys. Rev. E}
{\bf 60}, 7270 (1999).

\bibitem{lon98}
A.~Longtin, D.~Chialvo, \emph{Phys. Rev. Lett.} \textbf{81}, 4012
(1998); L.~Gammaitoni, P.~H\"{a}nggi, P.~Jung, and F.~Marchesoni,
\emph{Rev. Mod. Phys.} {\bf 70}, 254 (1998).

\bibitem{bul96}
A.R.~Bulsara and A.~Zador, \emph{Phys. Rev. E} {\bf 54}, R2185
(1996).

\bibitem{lev96}
J.E.~Levin and J.P.~Miller, \emph{Nature} {\bf 380}, 165 (1996).

\bibitem{sto01}
N.G. Stocks and R. Mannella, \emph{Phys. Rev. E} \textbf{64}, 030902
(2001); N.G.~Stocks, D.~Allingham, and R.P.~Morse, \emph{Fluctuation
and Noise Lett.} \textbf{2}, L169 (2002).

\bibitem{fit61}
R.~Fitzhugh, \emph{Biophys. J.} \textbf{1}, 445-466 (1961);
J.S.~Nagumo, S.~Arimoto, and S.~Yoshizawa, \emph{Proc. Inst. Radio
Engineers} \textbf{50}, 2061-2070 (1962).

\bibitem{pre93}
W.~Press, B.~Flannery, S.~Teukolsky, and W.~Vetterling {\em
Numerical Recipes in C}, Cambridge University Press, Cambridge,
1993.

\bibitem{pon33}
L.A.~Pontryagin, A.A.~Andronov, and A.A.~Vitt, \emph{Zh. Eksp. Teor.
Fiz.} \textbf{3}, 165 (1933).

\bibitem{gar04}
C. W. Gardiner, {\em Handbook of Stochastic Methods}, (Springer,
Berlin, 2004).

\bibitem{gar78} R.P. Garay and R. Lefever,
\emph{J. Theor. Biol.} \textbf{73}, 417 (1978); R. Lefever, W.
Horsthemke, \emph{Bull. of Math. Biol} \textbf{41}, 469 (1979).

\bibitem{och04}
A. Ochab-Marcinek and E. Gudowska-Nowak, \emph{Physica A}
\textbf{343}, 557 (2004).

\bibitem{sic04}
A. Mantovani, A. Sica, \emph{et al.}, \emph{Trends Immunol.}
\textbf{25}, 677 (2004); A. Mantovani, P. Allavena, A. Sica,
\emph{Eur. J. Cancer} \textbf{40}, 1660 (2004).

\bibitem{ell05}
R. L. Elliott and G. C.Blobe, \emph{J. Clin. Oncol.}
\textbf{23}, 2078 (2005).


\end{thebibliography}
\end{document}